# Morphology dependent decomposition and pore evolution during oxidation of Cr$_2$AlC coatings revealed by correlative tomography

**Devi Janani Ramesh[1*], Sameer Aman Salman[1], Jochen M. Schneider[1]**

[1] Materials Chemistry, RWTH Aachen University, Aachen, Germany

*Corresponding author: ramesh@mch.rwth-aachen.de

**Graphical Abstract**

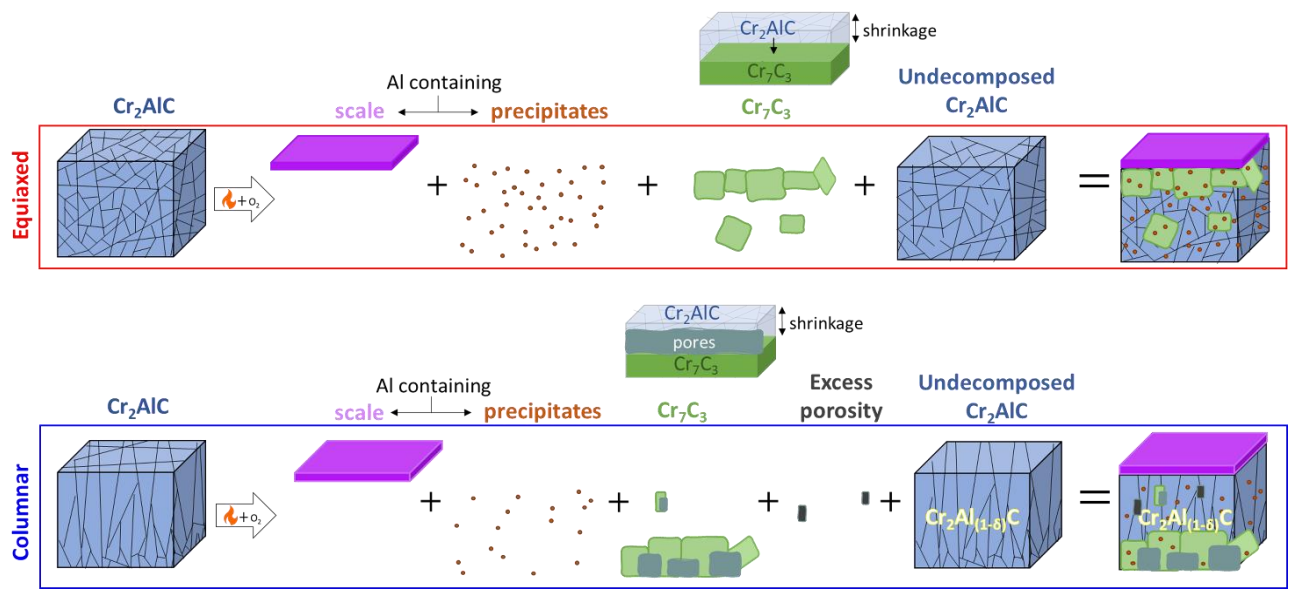

**Abstract**

Quantitative 3D characterization of materials degradation in oxidizing environments remains limited. Here, we apply a correlative tomography-based mass-balance framework to $Cr_2AlC$, a coating candidate for accident-tolerant nuclear fuel claddings and turbine blades, and show that decomposition and pore evolution during oxidation, quantified by integrating volumetric, structural and compositional data, are strongly governed by grain morphology.

The oxidation of sputtered $Cr_2AlC$ coatings with equiaxed and columnar grain morphologies was analyzed. While $Cr_7C_3$ formed in both coating morphologies, pores formed exclusively in columnar coatings. The expected $Cr_7C_3$ volume was estimated by mass-balance calculations assuming that Al-deintercalation enables oxide scale and Al-O-C-N precipitate formation, leading to complete transformation of the Al-deintercalated $Cr_2AlC$ into $Cr_7C_3$. In equiaxed coatings, the predicted carbide volume agreed with tomography within $3 \pm 3\,\%$, confirming Al-deintercalation-driven $Cr_7C_3$ formation. Despite the smaller molar volume of $Cr_7C_3$ relative to $Cr_2AlC$, absence of pores imply that transformation shrinkage is likely accommodated by coating thickness reduction. In columnar coatings, the predicted $Cr_7C_3$ volume exceeds the measured value by $22 \pm 4\,\%$, and the pore volume expected from transformation shrinkage alone is 13-16 % lower than measured, indicating partial Al deintercalation and clustering of pre-existing defects. This combined methodology provides a general route to quantitatively resolve degradation mechanisms.

## 1. Introduction

It is desirable to prevent degradation of materials during service to minimize economic losses as well as to ensure safety [1]. While regular inspection can help prevent major damage, the development of materials that are inherently more resistant to degradation is a more effective and long-term solution. The first step in realizing this is to understand how degradation occurs at the atomic, microscopic as well as macroscopic levels. Advances in characterization tools now allow for *in-situ* experiments that show how materials degradation initiates, and evolves in real time. However, it is not easy for all degradation processes to be studied this way, e.g., it is challenging to perform *in-situ* TEM studies maintaining realistic chemistry of the corrosive medium, and mitigating beam damage [2]. Furthermore, many *in-situ* experiments only provide information about the corrosion products without enabling complete characterization of the original material. For example, during *in-situ* XRD, the information depth is limited by the penetration depth of the X-rays, and lacks spatial resolution [3]. As a result, the most common and straightforward method to analyze material degradation is to examine materials after they have degraded, using sequential analysis to form hypotheses and trace back the degradation processes.

Although degradation clearly occurs in three dimensions (3D), most contemporary studies on the oxidation and corrosion of materials remain based on two-dimensional (2D) data. Although techniques such as computed tomography (CT) scans and atom probe tomography (APT) offer 3D insights, they excel but are simultaneously limited by their respective spatial resolution [4]. For example, while X-ray computed tomography (CT), offers access to larger volumes but with at lower spatial resolution while, atom probe tomography (APT) provides near-atomic resolution but is restricted to very small volumes [4]. These limitations are especially pronounced when analyzing coating materials, where coating thicknesses are on the order of microns and degradation product dimensions can span from a few tens of nanometers to several microns. Focused ion beam (FIB) tomography using serial sectioning uniquely allows 3D visualization of a material's microstructure while maintaining a spatial resolution of up to a few tens of nanometers, making it particularly well suited for studying degradation processes in coatings. In addition to visualizing the microstructure, this technique also provides the opportunity for quantitative analysis of various microstructural domains within the sample, provided they can be segmented by contrast differences [5–7]. The technique can be performed using either a backscatter electron detector or a secondary electron detector, which primarily provide compositional and topographic contrast, respectively. Furthermore, the serial sectional sectioning method can be combined with other analytical tools such as electron backscatter diffraction (EBSD) [8,9] and energy-dispersive X-ray spectroscopy (EDX) [9,10] to extract additional crystallographic and compositional information from the segmented regions, respectively. To date, FIB-SEM tomography has been widely used in fuel cell research for analyzing porosity and correlating the microstructure of the electrodes with synthesis conditions [11–13]. Some studies have also employed the method for analyzing the 3D electrode microstructure after operation [14,15]. Although FIB-SEM tomography has been employed in a few oxidation studies, it has primarily been used for visualization of microstructural features [16,17] and quantifying the porosity [16]. Nevertheless, its potential to comprehensively analyze microstructural evolution occurring throughout a process at different intervals remains largely untapped.

In this study, we study degradation phenomena occurring in $Cr_2AlC$ coatings using 3D data. $Cr_2AlC$ is a well-known MAX phase that has recently attracted considerable attention as a candidate coating for accident-tolerant fuel claddings in light water reactors, owing to its combination of high temperature oxidation resistance[18–22], and radiation tolerance[23]. The presence of Cr is expected to provide protection under normal operating conditions, while Al can form a passivating alumina scale at elevated temperatures, thereby mitigating catastrophic damage to the Zr cladding during loss of coolant scenarios[19,24]. It is also being explored as a candidate coating material for other high temperature applications, such as bond coats for thermal barrier coatings[25–27]. However, its oxidation behavior has been studied predominantly using 2D characterization techniques. [18,28–30].

In the case of bulk $Cr_2AlC$, oxidation has been shown to result in the formation of an alumina scale which proceeds by Al depletion from the bulk material and subsequently results in $Cr_7C_3$ formation [28,30–32]. During oxidation above approximately 1200 °C, pores have been reported to form in the $Cr_7C_3$ layer [28,33,34]. Zuber et al., oxidized samples at temperatures between 800 – 1400 °C, and reported pore formation in the $Cr_7C_3$ layer below the oxide scale, irrespective of the oxidation temperature [31]. These pores are suggested to result from the formation of gaseous carbon oxides and the redistribution of Al from the $Cr_2AlC$ to the oxide scale [28,31,32]. Recently, Reuban et al., reported the formation of pores in the $Cr_7C_3$ layer after oxidizing $Cr_2AlC$ for 40 minutes at 1000 °C [35]. They attributed the pore formation to volume shrinkage during the phase transformation of $Cr_2AlC$ to $Cr_7C_3$.

Studies on the oxidation behavior of $Cr_2AlC$ coatings have also reported similar $Cr_xC_y$ sub-layer formation [18,19,29]. Wang et al. observed the formation of tiny pores in the Cr-carbide layer beneath the oxide after oxidizing their equiaxed $Cr_2AlC$ coatings for 40 hours at 900 °C [18]. However, they did not report any pores at the coating-substrate interface despite the diffusion of Al into the Hastelloy substrates used. In a study by Michael et al., pore formation was detected in columnar $Cr_2AlC$ coatings near the Zr substrate and within the carbide regions beneath the oxide after oxidation for 15 minutes at 1100 °C in static air [19]. Similarly, Hajas et al. [29] and Chen et al. [36] reported pore formation of columnar $Cr_2AlC$ coatings within the Cr-carbide layer [29], and along grain boundaries [36], respectively, but provided no information on pore formation near the coating-substrate interface. The formation of pores is detrimental to the coating's performance, especially when formed at the coating-substrate interface as it mechanically weakens the coating. In these studies, pore formation was attributed to Al redistributed from $Cr_2AlC$ to the oxide scale [18,19,29] or to the growth of pores already present in the as-synthesized state during oxidation [36]. However, neither studies on the oxidation behavior of bulk $Cr_2AlC$ nor those on coatings have attempted to verify these hypotheses pertaining to oxidation induced decomposition and pore formation by mass balance calculations. Hence, the aim of this work is to assess if Al vacancy clustering driven pore formation occurs during oxidation of $Cr_2AlC$ and thereby understand the role of coating morphology on the degradation mechanism of $Cr_2AlC$. We develop and apply an approach using FIB-SEM tomography in combination with EDX in scanning transmission electron microscopy mode (STEM-EDX), and APT to compare theoretically predicted carbide volumes (from compositional/stoichiometric mass balance calculations) with experimentally measured volumes in $Cr_2AlC$ coatings with different initial morphologies - equiaxed and columnar. These two morphologies are chosen since they exhibit very different oxidation behavior, as detailed in our previous studies [36,37].

To this end, $Cr_2AlC$ coatings with columnar and equiaxed grain morphologies were synthesized using magnetron sputtering and then oxidized in ambient air up to 990 °C, and at 990 °C for durations of 2 and 3 hours. To achieve a 3D spatially resolved information on degradation of $Cr_2AlC$ coatings with the progress of oxidation, the decomposition of the MAX phase in the oxidized samples was investigated using a combination of correlative microscopy techniques, including FIB-tomography, STEM-EDX, and selected area electron diffraction (SAED). Mass balance calculations, based on volumetric and compositional data obtained through the correlative tomography approach were employed to identify the mechanisms underlying the oxidation induced decomposition and pore formation in $Cr_2AlC$ coatings of both morphologies.

## 2. Results and discussion

### 2.1. Composition and phase formation

Figure 1 presents cross-sectional HAADF-STEM images of as-synthesized and oxidized columnar and equiaxed coatings, while Figure 2 shows magnified HAADF-STEM images, TKD maps and reconstructed APT tips after transient oxidation. The compositions of the as-synthesized coatings were measured by a combination of elastic recoil detection analysis (ERDA) and electron backscatter spectroscopy (EBS), as described in [37]. The as-synthesized columnar coatings had an average composition of 49.8 ± 1.7 at.% Cr, 25.6 ± 0.9 at.% Al, 23.4 ± 0.8 at.% C, and 1.2 ± 0.1 at.% O, while the equiaxed coatings contained 49.9 ± 1.7 at.%

Cr, 23.8 ± 0.8 at.% Al, 23.7 ± 0.8 at.% C, 0.9 ± 0.03 at.% O, and 1.7 ± 0.1 at.% N. No compositional contrast is evident in the as-deposited coatings, see Figure 1 (a) and (e). On the other hand, oxidized coatings shown in Figure 1 (b)-(d), and (f)-(h) reveal a dark top layer corresponding to the oxide scale, bright regions within the coating are observed, indicating Cr-enriched regions embedded in a MAX phase matrix. Specifically in the oxidized columnar coatings, the presence of dark regions, indicating pores, is observed.

The oxide scale composition, and structure was investigated in our previous work [37]. Equiaxed coatings developed an oxide scale of $(Cr,Al)_2O_3$ with $α-Al_2O_3$ structure containing approximately 8.9 at.% Cr, under all oxidation conditions. In contrast, columnar coatings formed an oxide scale composed of a mixture of γ- and θ- $Al_2O_3$ with ≤ 1.4 at.% Cr, after transient oxidation, as indicated by selected area electron diffraction (SAED) and ERDA, which partially transformed to α- $Al_2O_3$ during isothermal oxidation [37]. We note here that the oxide scale composition measurement by ERDA discussed above was carried out on coatings synthesized on MgO substrates and oxidized to identical conditions. Oxide scale compositions measured by STEM-EDX for equiaxed coatings deposited on $α-Al_2O_3$ in the present study are consistent with ERDA measurements on equiaxed coatings deposited on MgO in [37]. However, for the columnar coatings, no Cr was measured by STEM-EDX while 1.4 at.% was measured in the columnar coatings deposited on MgO in [37].

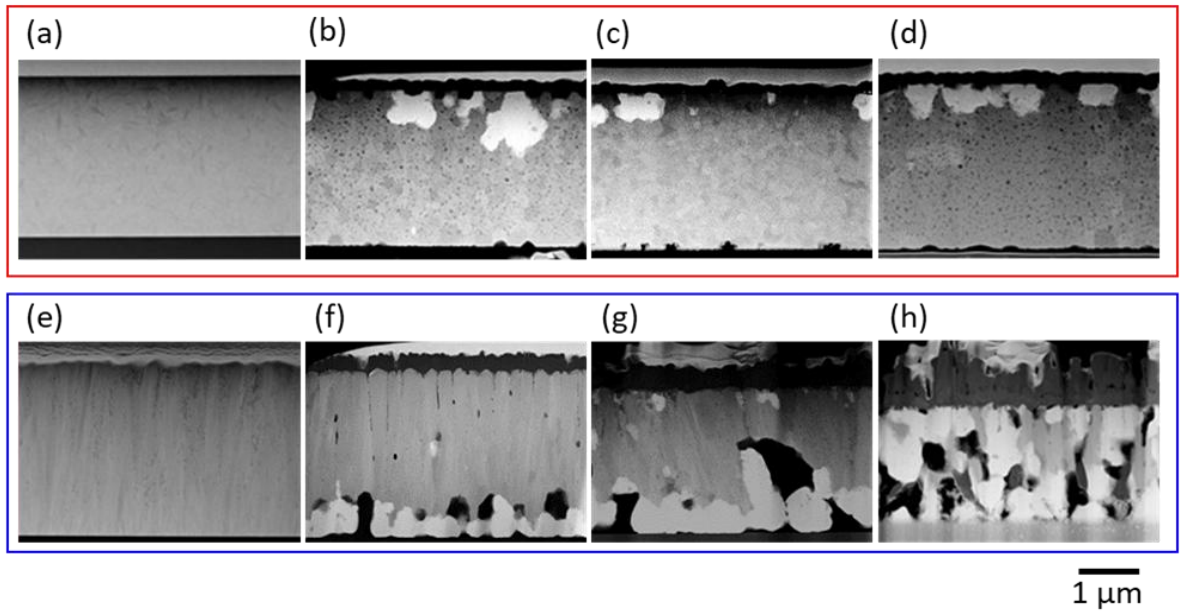

*Figure 1 HAADF - STEM images of (a), (e) – as-synthesized equiaxed and columnar coatings, respectively. (b) - (c) equiaxed coatings and (f) - (h) columnar coatings oxidized up to 990 °C, at 990 °C for 2 and 3 hours, respectively.*

The bright Cr-rich regions align with previous studies demonstrating that $Cr_2AlC$ oxidation causes decomposition and subsequent formation of Cr carbides - $Cr_7C_3$ and/or $Cr_3C_2$ [18,28,29,31,32]. Although some studies have employed SEM-EDX to determine the stoichiometry of Cr-carbides [28,31,32], the accuracy of SEM-EDX for the quantification of light elements such as carbon, is limited by the low X-ray energy and the low fluorescent yield resulting in a low signal-to-noise ratio and significant absorption before the X-rays can be detected [38]. Nevertheless, phase identification requires in addition to compositional data also structural information. To identify the structure of these carbides, Hajas et al. and Wang et al. performed SAED along the zone axis on different carbide regions [18,29]. The former showed the formation of $Cr_3C_2$ and $Cr_7C_3$ beneath the oxide scale in columnar coatings after oxidation at 950 °C and 1150 °C [29]. Wang et al. synthesized equiaxed coatings through a two-step process and observed the formation of $Cr_7C_3$ and $Cr_3C_2$ within the coating after oxidation at 900 °C for 40 hours [18]. These studies [18,29] confirm that both $Cr_7C_3$ and $Cr_3C_2$ can simultaneously form in columnar and equiaxed coatings during oxidation.

In the present work, the compositions of Cr-enriched regions were measured by STEM-EDX, which offers improved accuracy owing to the reduced information depth [39]. The measured Cr:C ratios ranged between 1.5 and 2.3, with a standard deviation of 5 at.%, and 4 at.% for Cr and C, respectively. Detailed analysis of phase formation and composition was carried out for coatings oxidized up to 990 °C; this condition was chosen for further analysis and will be discussed extensively in Section 2.3. After phase identification of the bright Cr-carbide regions was by SAED (see Figure S.1 in the Supplementary Data), TKD mapping was carried out, see Figure 2 (a) and (e). Although $Cr_7C_3$ was the only Cr-carbide phase identified by SAED (see Figure S3 in supplementary data), the TKD indexing was performed against $Cr_7C_3$, $Cr_3C_2$ (both Cnma and Pnma space groups), $Cr_{23}C_6$, and $Cr_2Al$. The TKD map see Figure 2 (a), reveals the presence of $Cr_2AlC$, $Cr_7C_3$ and small amounts of $Cr_2Al$ in equiaxed coatings, after transient oxidation. In contrast, only $Cr_2AlC$, and $Cr_7C_3$ were identified in columnar coatings, see Figure 2(e). This confirms that the brighter regions observed in both STEM-HAADF (Figure 1 (b) and (f)) are $Cr_7C_3$, with no $Cr_3C_2$ or $Cr_{23}C_6$ detected in the coatings after transient oxidation, irrespective of the initial grain morphology.

For equiaxed coatings, the Al concentration in the undecomposed MAX phase region of the coating, as measured by STEM-EDX was 27.5 ± 0.5 at.%, 27.6 ± 0.3 at.%, and 27.9 ± 0.9 at.% after transient oxidation and oxidation for 2 and 3 hours, respectively, compared to an average of 23.8 ± 0.8 at.% in the as-deposited condition. For columnar coatings, this was measured to be 26.4 ± 0.4 at.%, 26.1 ± 0.1 at.%, and 26.9 ± 0.2 at.% after transient oxidation and oxidation for 2 and 3 hours, respectively, versus an average of 25.6 ± 0.9 at.% in the as-deposited state. This increase in Al is attributed to the formation of Al-rich precipitates embedded in the undecomposed MAX phase region, as indicated in Figure 2 (b) and (f). Elemental distributions within these precipitates, and an accurate measurement of the increased Al content in the undecomposed MAX phase regions were obtained from APT analysis performed on specimens prepared from the undecomposed MAX regions, shown in Figure 2 (c), (d) and (g), (h). APT measurements were successfully conducted on equiaxed coatings oxidized up to 990 °C. As shown in Figure 2 (d), the proximity histogram reveals that the precipitate interior (corresponding to positive values along the x-axis) is Al rich and contains O, C, and N, and will be referred to as Al-O-C-N in the following text. Measurements on columnar coatings under the same oxidation conditions were unsuccessful, likely due to mechanical weakening of the APT tip due to pores within the coating (discussed in Section 2.2), which led to tip fracture either during preparation or at the onset of the measurement. Accordingly, for the columnar coating, an APT tip was prepared from a specimen oxidized to a maximum temperature of 850 °C. This tip captures an intermediate stage preceding complete precipitate formation as evidenced by a pronounced compositional gradient in the proximity histogram (see Figure 2 (h)) across the matrix-precipitate interface, as well as by the presence of Cr within the precipitate, which was absent in the precipitates forming in the equiaxed coatings (see Figure 2 (d)).

The proximity histograms shown in Figure 2 (d) and (h) were defined relative to iso-surfaces with lower Al concentrations (36 and 30 at.% for equiaxed and columnar, respectively) than the iso-surfaces used for tip reconstruction in Figure 2 (c) and (g) (40 and 33 at.% for equiaxed and columnar, respectively) to extend the compositional gradient region and to obtain an accurate matrix composition. The excess Al concentration in the undecomposed MAX phase region stemming from the Al-O-C-N precipitates was calculated by subtracting the matrix Al concentration obtained from the proximity histogram from the overall Al concentration of the analyzed tip. The uncertainty was determined from the standard deviation within the matrix region, as the counting errors are negligible. The resulting excess in Al content residing in precipitates in the undecomposed MAX phase was $2.7 \pm 0.2$ at.% in equiaxed coatings oxidized up to 990 °C and $0.9 \pm 0.1$ at.% in columnar coatings oxidized up to 850 °C. It is anticipated that the precipitate volume as well as the Al concentration within the precipitates in the undecomposed MAX phase in columnar coatings oxidized up to 990 °C will likely be larger, as observed in the STEM images, see Figure 2 (f). Hence the corresponding Al concentration in these precipitates may be underestimated here.

The oxygen and nitrogen contents measured by STEM-EDX in the coating remained similar between the as-synthesized and oxidized states, see Table S.1 in supplementary data. This suggests that the Al-O-C-N precipitates likely form as a result of aluminum reacting with oxygen and nitrogen impurities introduced during the sputtering process rather than an ingress of O or N from the atmosphere during oxidation. This observation is consistent with the propensity of aluminum to react with trace amounts of O and N to form thermodynamically stable oxide and nitride phases.

The Al-O-C-N precipitates identified in the $Cr_2AlC$ regions are also present in the carbide regions, see Figure 2 (b) and (f). As a consequence, STEM-EDX detected Al concentrations originating from Al-O-C-N precipitates within the carbide regions of 3.6 ± 0.4, 3.2 ± 0.2, and 3.4 ± 0.3 at.% in equiaxed coatings, and 1.8 ± 0.7, 1.2 ± 0.5, and 1.4 ± 0.1 at.% in columnar coatings after transient oxidation and oxidation for 2 and 3 hours at 990 °C, respectively. The higher content of Al due to precipitates within the carbide regions in equiaxed coatings compared to columnar coatings may be ascribed to higher impurity concentrations in the as-deposited state of the former.

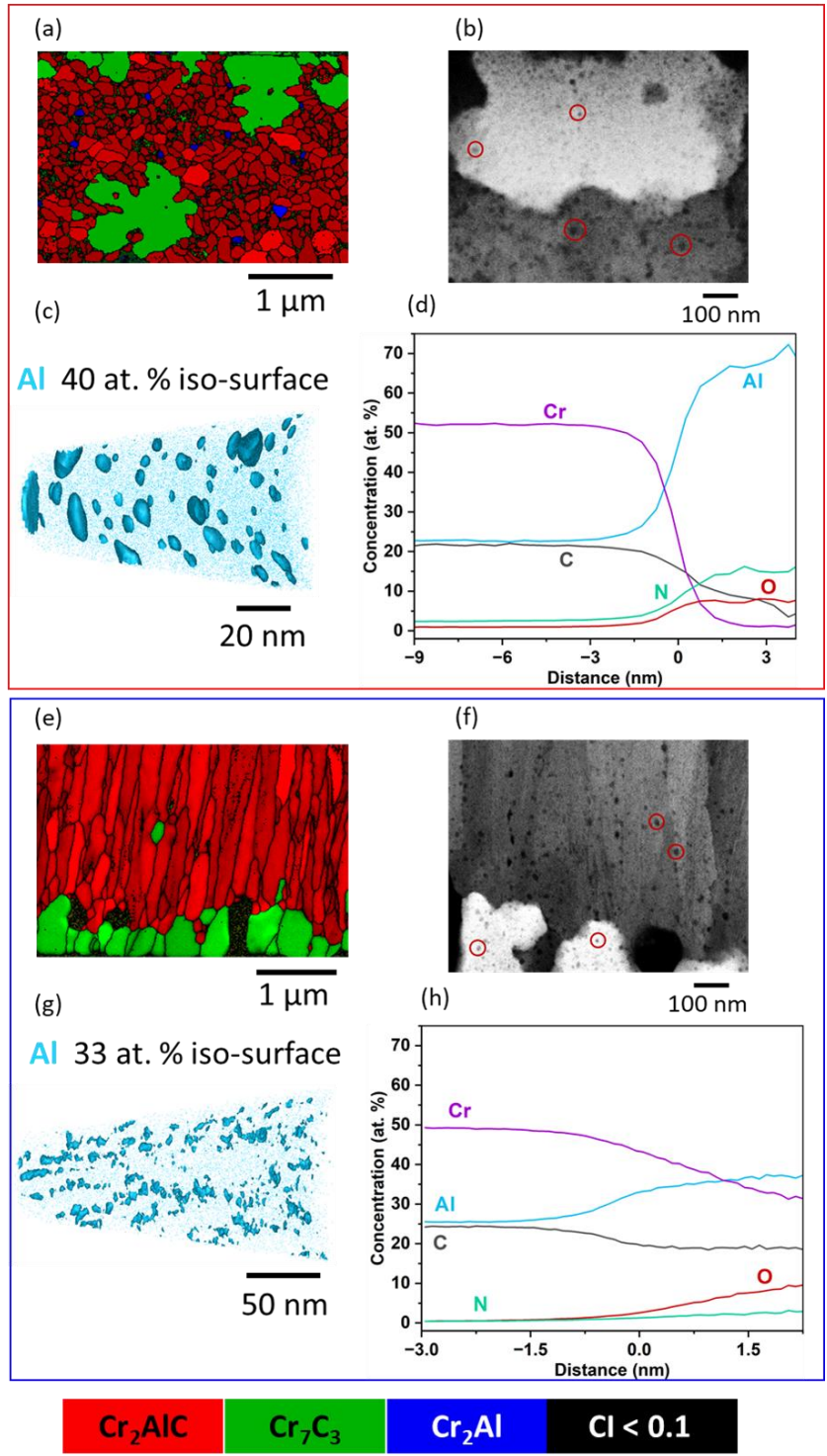

*Figure 2 (a) and (e) represent the phase maps of equiaxed coatings and columnar coatings oxidized up to 990 °C obtained by TKD. The maps are overlaid with the confidence index (CI) on grey scale, and the grain boundaries (misorientation of 15°) are included. Figures (b) and (f) are magnified HAADF-STEM images of equiaxed and columnar coatings oxidized up to 990 °C, respectively. The red circles in (b) and (f) indicate the presence of Al precipitates in these regions. Figure (c) shows a reconstructed APT tip with 40 at.% Al iso-surface, extracted from the undecomposed MAX phase region in equiaxed coating oxidized up to 990 °C. Figure (d) shows the proximity histogram from the tip in (c) at interfaces with 36 at.% Al isosurface – distance 0 to 3.5 nm shows the composition into the Al rich precipitates (> 36 at.%), while 0 to -9 nm shows the composition into the $Cr_2AlC$ region. Figure (g) shows a reconstructed APT tip with 33 at.% Al iso-surface, extracted from the undecomposed MAX phase region in columnar coating oxidized up to 850 °C. Figure (h) shows the proximity histogram from the tip in (g) at interfaces with 30 at.% Al isosurface– distance 0 to 2.2 nm shows the composition into the Al rich (> 30 at.%) precipitates, while 0 to -3 nm shows the composition into the $Cr_2AlC$ region.*

## 2.2. SEM tomography

A representative SEM tomography volume, measuring 20 μm × (coating thickness + scale thickness) μm × 2 μm in depth, is shown in Figure 3. The data were first segmented and then reconstructed from four distinct data sets collected for each grain morphology from the as-synthesized and the oxidized coatings. These reconstructions serve to aid visualization and, together with a larger volume dataset, were employed for the quantitative analysis presented in Section 2.3. As demonstrated therein, the smaller reconstructed volume accurately represents the decomposition behavior. The white regions within the coating in Figure 3 represent the undecomposed MAX phase. For reference, tomography of the as-synthesized coatings was also performed, and is presented in Figure 3 (a) and (e), for equiaxed and columnar coatings, respectively. Nano-pores along the column boundaries close to the surface of the coating, also seen in plan-view STEM images [37], could be observed in the columnar coatings in the as-synthesized conditions, while no compositional or morphological contrast could be observed in the equiaxed coatings, see Figure S.2 in supplementary data. The contrast observed in the SEM images, Figure 3 are identical to that in the STEM-HAADF images, see Figure 1. This confirms that the bright regions identified in the SEM are $Cr_7C_3$. In equiaxed coatings after transient oxidation, trace amounts of $Cr_2Al$, see Figure 2 (a), were identified in addition to $Cr_7C_3$. These regions show very little contrast variation relative to $Cr_2AlC$ which is consistent with [40]. Consequently, $Cr_2AlC$ and $Cr_2Al$ were grouped together during tomography.

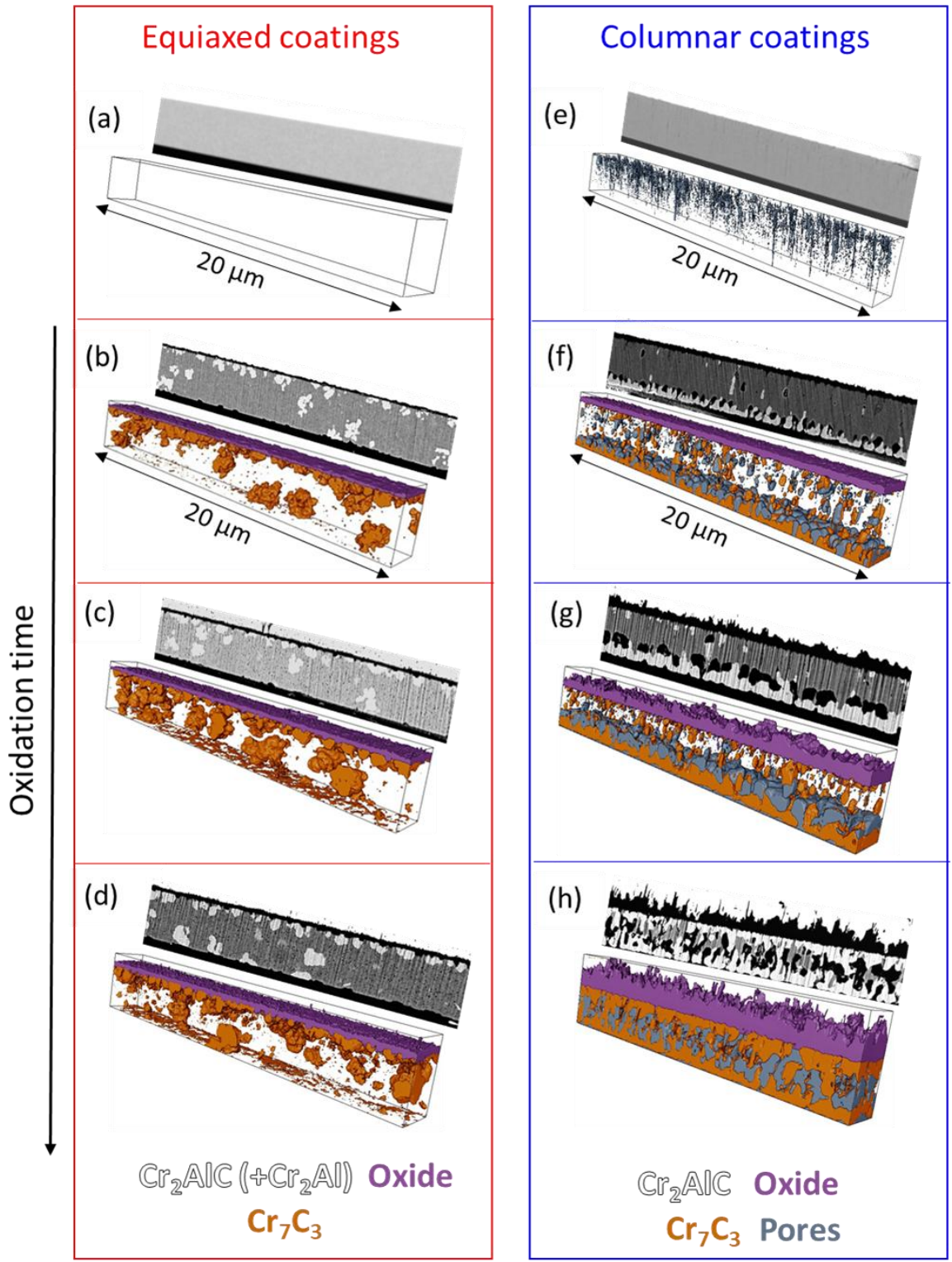

*Figure 3 Each panel represents a representative SEM image, at the top, and the reconstructed SEM tomography obtained from a stack of imilar SEM images at the bottom. (a)-(d), represent equiaxed coatings and (e)-(h) represents columnar coatings at different conditions. (a) and (e) represent the as-synthesized condition. (b) and (f) represent the condition after oxidation up to 990 °C. (c) and (g) represent the condition after oxidation for 2 h at 990 °C. (d) and (h) represent the condition after oxidation for 3 h at 990 °C.*

Figure 4 shows the average oxide scale thickness as well as carbide and pore volume fractions in the equiaxed and columnar coatings oxidized at different conditions.

The average oxide scale thicknesses, see Figure 4 (a), were determined by dividing the volumes of the oxides by their surface area, with both quantities determined from SEM tomography. The average oxide scale thickness in equiaxed coatings increases parabolically from 126 nm after transient oxidation to 215 and 236 nm after isothermal oxidation for 2 and 3 h at 990 °C, respectively. This positively correlates with the measured carbide volume fraction, see Figure 4 (b). In equiaxed coatings, the carbide volume fraction showed a slight initial increase from 0.11 after transient condition to 0.17 after 2 hours of isothermal oxidation and then remained almost constant, with a slight increase to 0.18 after 3 hours of isothermal oxidation. On the other hand, the increase in average oxide scale thickness is linear in columnar coatings, see Figure 4 (a). The average oxide scale thickness increased from 270 nm after transient oxidation to 520 and 872 nm after 2 and 3 h of isothermal oxidation at 990 °C, respectively. This rapid increase in scale growth positively correlates with the increase in both carbide and pore volume fractions in columnar coatings, see Figure 4 (b) and (c). The carbide volume fraction increases from 0.1 after transient oxidation to 0.24 and 0.64 after isothermal oxidation for 2 and 3 h at 990 °C, respectively. The pore volume fractions increases from 0.005 in the as-synthesized state to 0.088 after transient oxidation, followed by an increase to 0.159, and 0.271 after isothermal oxidation for 2 and 3 h at 990 °C, respectively. The pores after oxidation in columnar coatings are present at the coating-substrate interface surrounded by carbides, as well as along the coating, see Figure 3 (f) to (h). No evidence of pore formation was observed in equiaxed coatings at any oxidation condition.

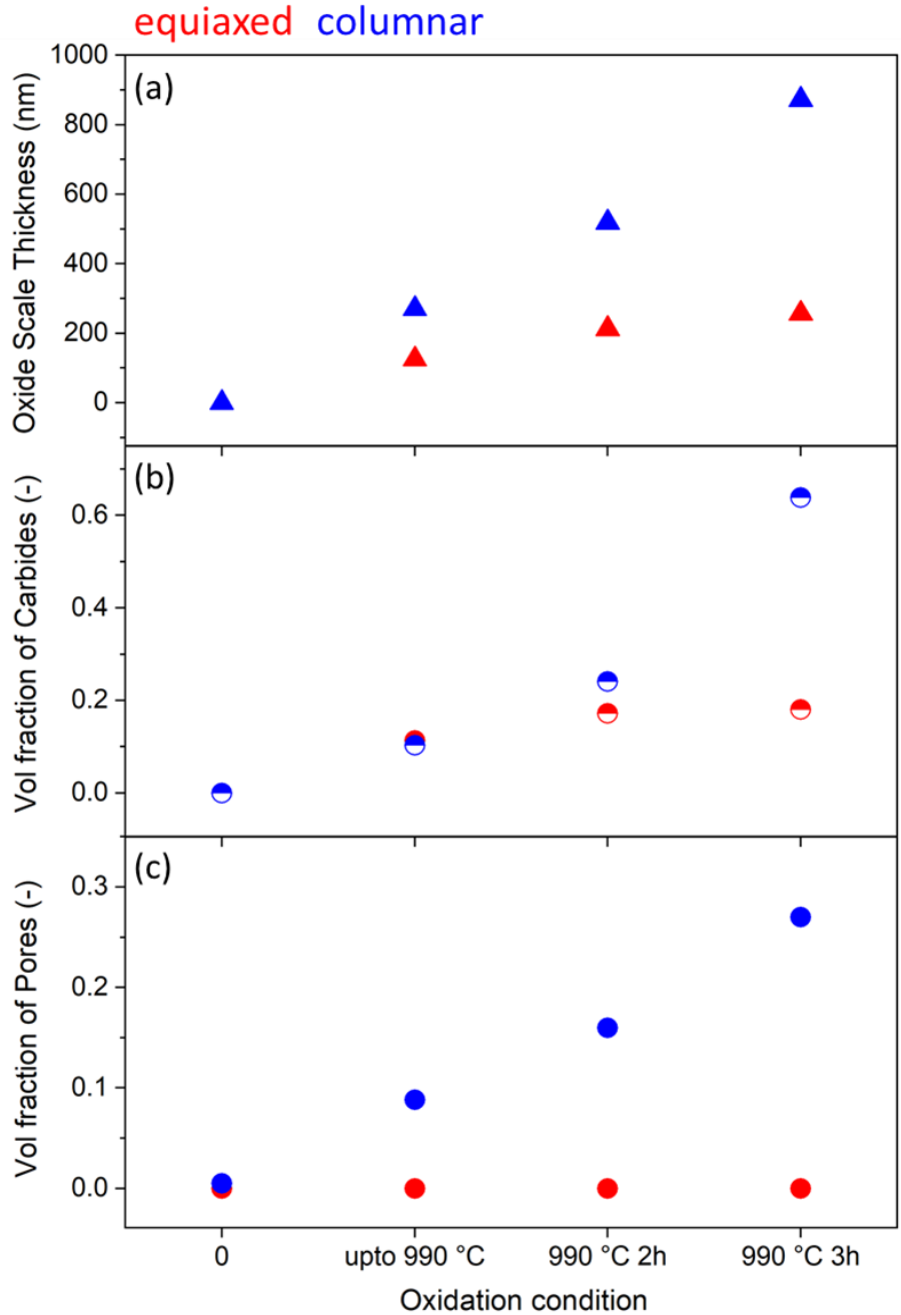

*Figure 4 (a) Oxide scale thickness formed at different oxidation conditions in equiaxed and columnar coatings obtained from SEM tomography at different oxidation conditions (b) volume fraction of carbides in the coating at different oxidation conditions obtained from SEM tomography. (c) volume fraction of pores in the coating at different oxidation conditions obtained from SEM tomography.*

2.3. Mass-balance calculations and comparison of the model with experiment

During oxidation of $Cr_2AlC$, mobile Al atoms are redistributed to the oxide scale (see Figure 1) as well as to the Al-O-C-N precipitates (see Figure 2). Upon complete deintercalation of Al from the MAX phase, the structure collapses into metastable $Cr_2C$ [41], which then transforms $Cr_7C_3$, as evidenced by TKD observations (Figure 2a and 2e).

Combining volumetric data obtained from tomography with local compositional data from STEM-EDX and APT, and applying mass-balance calculations under a defined set of assumptions, allows for the estimation of the theoretical volume of $Cr_7C_3$ that would form as a consequence of Al incorporation into the scale and the Al-O-C-N precipitates. Comparison of this predicted volume with the $Cr_7C_3$ volume measured by SEM tomography, provides a means of validating both the analysis and the underlying assumptions.

The mass balance calculations presented here are based on the following assumptions: (i) Al deintercalation enables oxide scale and Al-O-C-N precipitate formation, leading to a complete transformation of the Al-depleted $Cr_2AlC$ into $Cr_7C_3$ with no partially deintercalated MAX phase remaining; (ii) all phases exhibit theoretical densities at room temperature; and (iii) oxide scale formation prevents Al volatilization, ensuring that all deintercalated Al is retained.

A schematic model to predict the theoretical volume of $Cr_7C_3$ that should form based on the above assumptions is illustrated below in Figure 5. Details of the calculation procedure incorporating the compositional data obtained by STEM-EDX/APT as well as oxide and MAX phase volumes obtained by SEM tomography are provided in Section 4.3, while spreadsheets containing the raw data used are available as additional supplementary files.

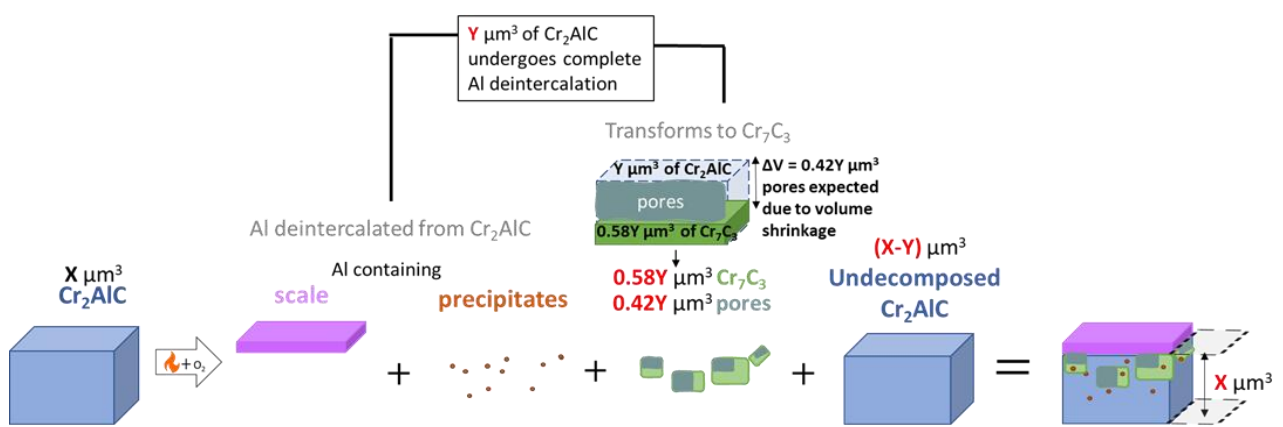

*Figure 5 Schematic model of the expected outcome during oxidation, based on volume estimates from mass-balance calculations.*

Table 1 presents the presents predicted and SEM tomography-measured carbide volumes in equiaxed and columnar coatings. To ensure repeatability and assess the accuracy of the SEM tomography reconstruction, mass-balance calculations were performed on two independent volume segments of different sizes, smaller ($^s$) and larger ($^l$), see Table 1, extracted from the same samples. In equiaxed coatings, the predicted carbide volumes are 0 - 6% lower than the measured values, whereas in columnar coatings the estimated volumes exceed the measured values by 16 - 28%. Although some error may arise due to deviation of actual densities to the theoretical values assumed, these uncertainties are difficult to quantify; therefore, the uncertainties reported in Table 1 were derived solely from the compositional measurements of the excess Al and the Al content in the carbide. The deviations in both compositional measurements alter the overall result in the same direction, hence, the error bar represents the sum of uncertainties from these measurements. Additionally, comparing results from the two different reconstructed volumes suggests that the statistical error introduced by the tomography technique falls within the margin established by the compositional analysis. Notably, the smaller reconstructed volume is already representative of the sample. Furthermore, the methodologies for mass balance and tomography were consistently applied to both coating morphologies, indicating that any significant deviations between the estimated and measured carbide volumes are likely due to sample-specific mechanisms rather than the measurement techniques themselves.

Table 1 The table shows the carbide volumes measured from tomography ($V_{Cr_7C_3}^{meas}$), carbide volume predicted from the mass-balance calculations ($V_{Cr_7C_3}^{pred}$) after oxidation up to 990 °C. $^s$ and $^l$ correspond to different reconstructed volumes from the same sample. The estimated carbide volume is most sensitive to the excess Al content measured in Cr$_2$AlC; hence the error bars here represent the uncertainty in the estimated carbide volume, based on the uncertainty of this composition measurement.

| Quantity | Sample | | | |
|---|---|---|---|---|
| | Equiaxed$^s$ | Equiaxed$^l$ | Columnar$^s$ | Columnar$^l$ |
| $V_{Cr_7C_3}^{meas}$ / μm³ | 13.5 | 48.6 | 13.1 | 48.1 |
| $V_{Cr_7C_3}^{pred}$ / μm³ | 13.1 | 48.0 | 15.8 | 58.7 |
| $\dfrac{V_{Cr_7C_3}^{pred} - V_{Cr_7C_3}^{meas}}{V_{Cr_7C_3}^{meas}} \cdot 100$ | -3 ± 3 % | -1 ± 4 % | 20 ± 4 % | 22 ± 4 % |

The within-error agreement of -3 ±3 % between predicted and measured volumes in equiaxed coatings strongly supports our assumption; the notion that Al diffusion leading to oxide scale and Al-O-C-N precipitate formation is accompanied by complete Al deintercalation from Cr$_2$AlC, followed by its phase transformation to Cr$_7$C$_3$.

The phase transformation from Cr$_2$AlC to Cr$_7$C$_3$ entails substantial volume shrinkage (42%), which has previously been considered only by Reuban *et al.*[35] in their analysis of oxidized bulk Cr$_2$AlC, where they attributed pore formation to this shrinkage[35]. In the present work, however, no pore formation was observed in equiaxed coatings by either SEM or STEM under any investigated oxidation condition, see Figure 3 and Figure 1. This implies that the transformation induced shrinkage is accommodated differently, most plausibly by a reduction in coating thickness[42]. Figure 6 presents a schematic of the expected outcome from mass-balance calculations, compared with the experimentally observed mechanism during oxidation in equiaxed coatings.

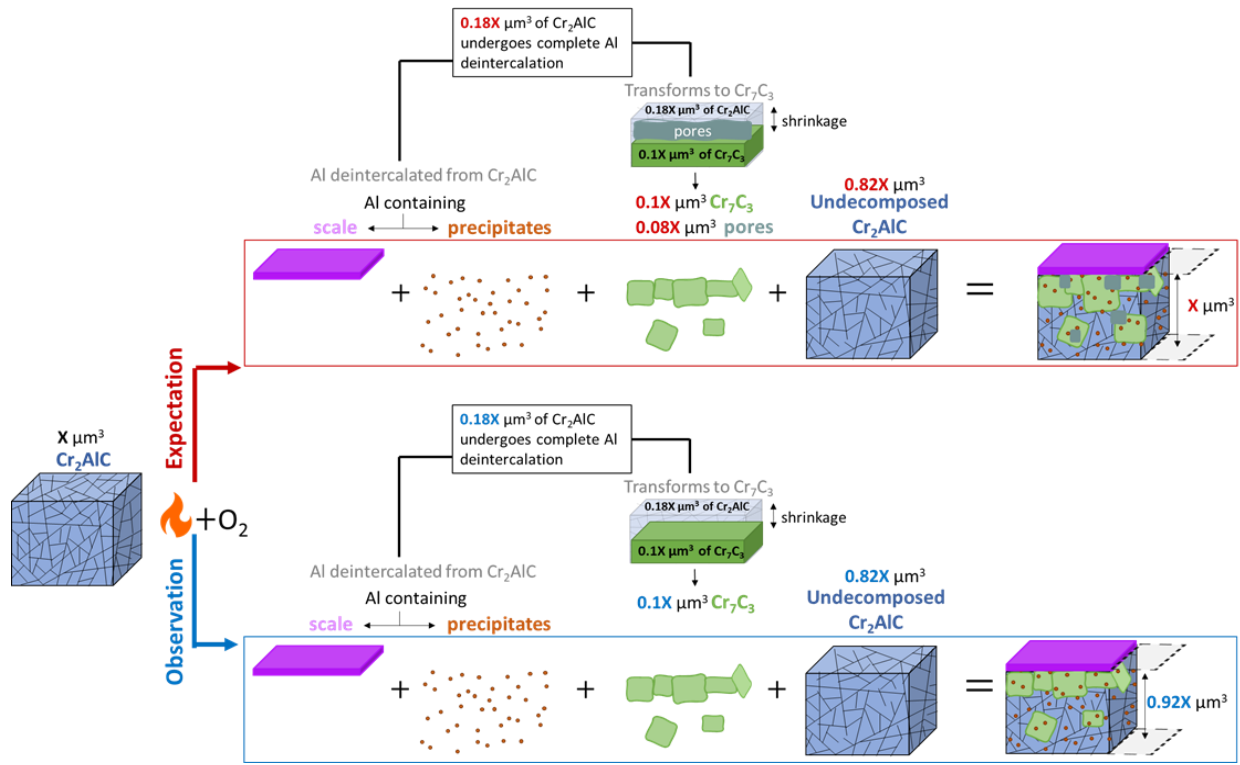

*Figure 6 Schematic representation of the expected outcome from mass-balance calculations, compared with the experimentally observed mechanism during oxidation in equiaxed coatings.*

In columnar coatings, the $Cr_7C_3$ volume predicted from mass-balance calculations is larger than the measured volume by 22 ± 4%. This deviation indicates that the Al concentration measured in the precipitates and oxide scale is lower than that inferred from the measured volume of $Cr_7C_3$, which is formed by complete deintercalation of Al from $Cr_2AlC$. While the error of ± 4 % is based on the statistical error associated with Al concentration measurement in the precipitates, the systematic error contributions associated with the Al concentration determination in the MAX phase, oxide scale and precipitates will increase the magnitude of the error. However, the quantification of excess Al present as Al-O-C-N in $Cr_2AlC$ is likely underestimated in columnar coatings, since it is derived from APT measurements taken at an earlier oxidation stage. This underestimation likely reduces the apparent deviation between the predicted and measured $Cr_7C_3$ volumes, so the true discrepancy might be larger. Hence, it is reasonable to assume that the 22 ± 4 % discrepancy arises from a combination of measurement uncertainties, such as an overestimation of Al deintercalation, or from limitations of the model which originate from the underlying assumption that only carbide formation caused by complete deintercalation of Al from the MAX phase occurs. When partial deintercalation of Al is considered, a similar Al content in the oxides and precipitates may cause a reduced carbide volume since Al vacancies resulting from partial deintercalation do not contribute to carbide formation. Their configurations at this point in time cannot be resolved by SEM tomography, and are subject to future investigations with suitable methods. However, the resulting reduction in Al concentration due to the partially deintercalated Al present within the unreacted $Cr_2AlC$ is calculated to be only ~0.8 - 0.9 at.%, well within the reported stability range of the MAX phase [35], indicating that such a mechanism is plausible. The formation of $Cr_7C_3$ as the sole carbide phase, necessitates a balance of the oxidation reaction through the generation of $CO/CO_2$ by-products, that have been associated with pore formation in earlier studies [28,31,43]. However, the formation of such large pores, as observed in the columnar coatings, see Figure 3 (e) – (h), cannot be solely attributed to $CO/CO_2$ formation, as also noted in [31]. Furthermore, the presence of Al-O-C-N precipitates, which incorporate some of the carbon released during $Cr_7C_3$ formation, and the potential formation of Cr vacancies in $Cr_7C_3$ that shift the Cr:C ratio closer to that of the MAX phase, may offer alternative pathways for balancing the C concentrations. Table 2 compares the pore volumes measured by SEM tomography, and the pores expected from shrinkage due to $Cr_7C_3$ formation in the two reconstructed volumes of columnar coatings. The porosity in columnar coatings increases substantially from ~0.5% volume fraction in the as-synthesized state to ~ 8.8% after oxidation up to 990 °C, see Figure 3. This is consistent with volume shrinkage associated with the $Cr_7C_3$ transformation, supporting Reuban *et al.*'s observations in bulk samples [26]. However, the estimated pore volumes expected solely from shrinkage, are 13–16% lower than those measured (after subtracting the baseline porosity of the as-synthesized state, resolvable with SEM tomography) see Table 2.

*Table 2 Comparison of pore volumes measured by SEM tomography, pre-existing pores in the as-synthesized state, and pores expected from shrinkage in the two reconstructed volumes.*

| Sample | Measured volume of pores ($\mu m^3$) | Volume of pores from as-synthesized state ($\mu m^3$) | Volume of pores expected due to formation of $Cr_7C_3$ ($\mu m^3$) |
|---|---|---|---|
| Columnar[1] | 11.3 | 0.5 | 9.0 |
| Columnar[2] | 39.8 | 1.7 | 33.0 |

It is reasonable to assume that the different synthesis strategies utilized to grow columnar and equiaxed coatings, result in different defect structures. Columnar coatings, produced by magnetron sputtering, inherently contain ion bombardment induced point defects such as vacancies [44], whereas equiaxed coatings undergo vacuum annealing from the amorphous state, a process that facilitates defect annihilation during crystallization. An earlier study reported elongated Cr-Al bond lengths near nanopores in columnar coatings synthesized by magnetron sputtering [45]. Hence, it is reasonable to assume that complex defect structures are formed during plasma assisted $Cr_2AlC$ coating synthesis. The estimates of pore volume presented here clearly indicate that pore formation in columnar coatings is primarily driven by shrinkage associated with $Cr_7C_3$ formation, while the additional pore volume unaccounted for by shrinkage likely arises from a combination of clustering of pre-existing nanopores and of intrinsic vacancies. Figure 7 presents a schematic of the expected outcome from mass-balance calculations, compared with the experimentally observed mechanism during oxidation in columnar coatings.

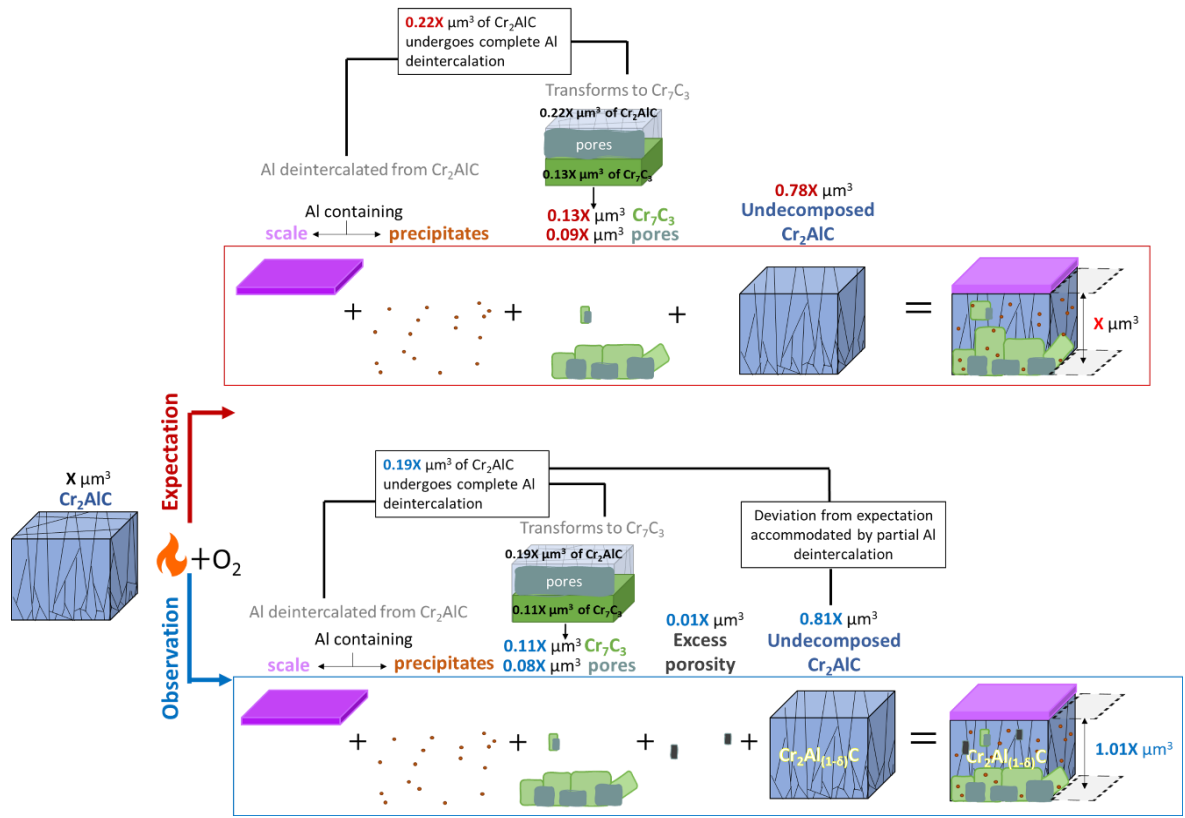

*Figure 7 Schematic representation of the expected outcome from mass-balance calculations, compared with the experimentally observed mechanism during oxidation in columnar coatings.*

## 3. Conclusions

The decomposition and pore formation in $Cr_2AlC$ coatings with equiaxed and columnar grain morphologies during oxidation up to 990 °C was investigated using a correlative approach combining FIB-SEM tomography for volume quantification, STEM-EDX, and APT for chemical composition, and TKD for structural information. During oxidation, Al deintercalation from the $Cr_2AlC$ led to the formation of oxide scale and Al-O-C-N precipitates dispersed in the undecomposed $Cr_2AlC$. Al de-intercalated regions subsequently transformed to $Cr_7C_3$, the latter also containing Al-O-C-N precipitates.

For equiaxed coatings, the mass-balance calculations showed within-error agreement of 3 ± 3%, between predicted and measured carbide volumes, validating the notion that Al deintercalation is complete and manifests solely as decomposition to $Cr_7C_3$. Despite the large volume shrinkage associated with this transformation, pores did not form; instead, shrinkage was likely accommodated by a reduction in coating thickness. In contrast, the predicted $Cr_7C_3$ volume exceeded the measured value by 22 ± 4% in columnar coatings. This sizable discrepancy invalidates the assumption that only complete Al deintercalation resulting in $Cr_7C_3$ formation occurs, and suggests that partial deintercalation of Al occurs resulting in Al vacancy formation in $Cr_2AlC$. Here, transformation shrinkage is accommodated primarily by pores, with porosity increasing from ~0.5% in the as-synthesized state to ~8.8% after oxidation up to 990 °C. However, the estimated pore volume expected from shrinkage alone is 13 - 16% lower than the measured pore volume. This excess measured porosity is rationalized by the clustering of pre-existing defects inherent to magnetron sputtered coatings.

These findings demonstrate the mechanisms leading to the superior oxidation resistance of equiaxed coatings, with complete Al deintercalation forming $Cr_7C_3$ and shrinkage accommodated without pore formation. In contrast, columnar coatings form extensive pores due to volume shrinkage occurring during the phase transformation to form $Cr_7C_3$, in addition to the clustering of pre-existing nano-pores and vacancies. This mechanistic difference provides a framework for tailoring MAX-phase microstructures for enhanced durability, while the correlative methodology developed here offers a general strategy to link local chemical transformations with large-scale structural evolution in complex coating and bulk systems.

## 4. Materials and methods

The coatings analyzed in this work were synthesized using an industrial-scale magnetron sputtering chamber (CC800/9, CemeCon AG, Wuerselen, Germany) on inert single crystal 10 × 10 × 0.5 mm α-$Al_2O_3$ substrates to prevent interactions at the coating-substrate interface.

Crystalline columnar $Cr_2AlC$ textured such that (110) planes were parallel to the substrate surface were produced by sputtering a 500 × 88 $mm^2$ compound $Cr_2AlC$ target (Plansee Composite Materials GmbH, Germany) in direct current mode at a power density of 2.27 Wcm-2 under a base pressure of <$10^{-3}$ Pa, with Ar as the working gas at a flow rate of 200 sccm. The substrates were externally heated to a temperature of 580 °C prior to starting the deposition.

To obtain equiaxed grain morphology, amorphous coatings were synthesized under identical parameters with the exception of intentional substrate heating. These coatings were subsequently annealed at 700 °C in a vacuum environment (base pressure < 5 × $10^{-4}$ Pa) for 2 hours resulting in the formation of crystalline $Cr_2AlC$ without pronounced texture. Both coatings were phase-pure as confirmed by X-ray diffraction (XRD). Further details on synthesis parameters, composition measurement, morphological and structural analysis of the as-synthesized coatings can be found in our previous study [37].

The synthesized coatings were oxidized in ambient air up to 990 °C at a heating rate of 33 K/min (transient oxidation), followed by isothermal oxidation at 990 °C for 2 and 3 hours.

4.1. Morphological, compositional and structural analysis of oxidized coatings

The microstructural evolution of the coatings - including morphology, composition, phase formation, and the volume fractions of different phases within the coating was analyzed using a combination of high angle annular dark field STEM (HAADF-STEM), STEM-EDX, SAED, and focused ion beam (FIB) tomography, respectively. SAED was performed on samples subjected to transient oxidation conditions, while HAADF-STEM, STEM-EDX, and FIB tomography was performed on samples oxidized to transient as well as isothermal conditions.

For TEM and APT analysis, thin lamellae and tips, respectively, were prepared using the Helios NanoLab660 system (FEI, Hillsborough, USA). (S)TEM imaging and selected area electron diffraction (SAED) was performed using a JEOL JEM-F200 (JEOL, Tokyo, Japan) microscope operating at an acceleration voltage of 200 kV. EDX measurements were performed using an Oxford UltimTE detector (Oxford instruments, Abingdon, United Kingdom) installed on the same instrument. The composition of various regions was determined using the Cliff–Lorimer method with theoretical k-factors. Absorption corrections were applied within the acquisition software *(AZtec)* by inputting the theoretical densities of the phases and the approximate lamella thickness measured after thinning. The standard deviation from measurements of the same phase in different regions along the lamella was used as the error bar, as the count rate error was significantly smaller than this variability.

Helios 5 Hydra UX dual-beam microscope (ThermoFisher Scientific, Waltham, USA) was used to record the transmission Kikuchi diffraction (TKD) maps at an acceleration voltage of 30 kV and a beam current of 6.4 nA at a step size of 9 nm. Kikuchi patterns were captured using the EDAX Velocity detector, and the indexing was carried out by the spherical indexing method. The analysis was carried out in EDAX OIM 9.0 Analysis software.

APT measurements were performed on specimens extracted from Equiaxed and columnar coatings oxidized up to 990 °C, and 850 °C, respectively. The measurement was carried out using a CAMECA local electrode atom probe (LEAP) 4000X HR, assisted by thermal pulsing with a laser energy 30 pJ, and at a frequency of 125 kHz and 200 kHz, for the equiaxed and columnar samples, respectively. The average detection rate was 0.5%, while the base temperature was maintained at 60 K.

4.2. FIB tomography - image acquisition and reconstruction

FIB tomography of both as-synthesized and oxidized coatings was performed using a Helios 5 Hydra UX microscope (Thermo Fisher Scientific, Oregon, USA) equipped with a multi-ion plasma source (argon, nitrogen, xenon, and oxygen). A 25 × 25 x ~3 µm chunk, was extracted from the as-synthesized and oxidized samples. The chunk was attached to the edge of a silicon wafer on a holder tilted at 54°, allowing parallel milling by tilting the holder to 16° (with the Pt surface perpendicular to the ion column) and imaging by tilting to 54° (with the cross-sectional surface of the coating perpendicular to the electron column). A fiducial marker was created near the Pt protective layer to correct for any milling position drifts caused by stage movements between milling and imaging.

Automated milling and imaging were performed using the software *Auto Slice and View 5.8* (Thermo Fisher Scientific, Oregon, USA). A $Xe^+$ beam accelerated at 30 kV and 740 pA was used to mill 20 nm thick slices before each SEM image capture. Ion beam images were taken at high enough magnification and resolution to accurately position the 20 nm milling window. A rocking mill with a 3° tilt was employed to reduce curtaining during milling. SEM images were captured using a concentric backscatter detector to achieve compositional contrast, with inner rings selected to enhance this contrast by detecting high-angle backscattered electrons (BSEs).

The optimal acceleration voltage was determined through Monte Carlo simulations using the software *Casino* [46], specifically to accurately quantify and reconstruct Cr carbide phases [47,48]. The simulations indicated that 5 kV was the optimal acceleration voltage for this material system, see Figure S.3 in supplementary data, given that the slicing thickness is 20 nm. Images of as-synthesized samples were acquired with a current of 0.2 nA and acceleration voltages of 3 kV, while the latter was set to 5 kV for the carbide-containing oxidized samples.

Images were captured with a 3 µs dwell time, and a voxel size of 5 × 5 × 20 nm³. The smaller x-y voxel dimensions were chosen to preserve the sharp edges of the features. Over 100 images from each tomography, covering depths of ≥ 2 µm, were used for reconstruction. Prior to reconstruction, a median filter with a 9×9 kernel was applied to reduce noise and achieve a grayscale distribution with discrete peaks for accurate segmentation. Reconstruction was performed using the software *Avizo* (Thermo Fisher Scientific, Oregon, USA), where consecutive slices were aligned using the least squares method and a uniform slice thickness of 20 nm was maintained. Segmentation was then carried out by thresholding regions based on grayscale values, as illustrated in Figure S.4, in supplementary data.

### 4.3. Mass-balance calculations

Mass-balance calculations, by taking inputs from the FIB-SEM tomography, TKD, and APT, were performed to predict the volume of $Cr_7C_3$ formed during the oxidation of the $Cr_2AlC$. The mass balance calculations presented here are based on the following assumptions: (i) Al deintercalation enables oxide scale and Al-O-C-N precipitate formation, leading to a complete transformation of the Al-depleted $Cr_2AlC$ into $Cr_7C_3$ with no partially deintercalated MAX phase remaining; (ii) all phases exhibit theoretical densities at room temperature; and (iii) oxide scale formation prevents Al volatilization, ensuring that all deintercalated Al is retained.

The calculations are summarized below, with the numerical values, and the spreadsheet used for the calculations is provided as supplementary data. The densities of the different phases used for the calculations were obtained from the following ICSD record numbers: $Cr_2AlC$ - 42918, $Cr_7C_3$ - 52289, $Cr_3C_2$ - 52289, α-$Al_2O_3$ - 52648, γ-$Al_2O_3$ - 66559, $Cr_2O_3$ – 167278. [49–54]

The number of moles of each atomic species $i$ present in phase $j$ is denoted by $n_i^j$. The number of moles of Al in the Al-O-C-N precipitates occurring in $Cr_2AlC$ and $Cr_7C_3$ post-oxidation are denoted respectively by $n_{Al}^{ppt,Cr_2AlC}$ and $n_{Al}^{ppt,Cr_7C_3}$.

Pre-oxidation, all Cr and Al atoms are present in $Cr_2AlC$.

$$n_{Cr}^{tot,b.o} = n_{Cr}^{Cr_2AlC,b.o} \tag{1.1}$$

$$n_{Al}^{tot,b.o} = n_{Al}^{Cr_2AlC,b.o} \tag{1.2}$$

$n_i^{tot,b.o}$ and $n_i^{Cr_2AlC,b.o}$ respectively denote the total moles of $i$ atoms and moles of $i$ atoms in the $Cr_2AlC$ phase present before oxidation.

As both the columnar and equiaxed coatings consist of stoichiometric phase-pure $Cr_2AlC$ prior to oxidation, we consider

$$\frac{n_{Cr}^{Cr_2AlC,b.o}}{n_{Al}^{Cr_2AlC,b.o}} = 2 \tag{1.3}$$

The total number of Cr and Al moles after oxidation, $n_{Cr}^{tot,a.o}$ and $n_{Al}^{tot,a.o}$, respectively are given by Equations (2.1) and (2.2).

$$n_{Cr}^{tot,a.o} = n_{Cr}^{Cr_2AlC,a.o} + n_{Cr}^{oxide} + n_{Cr}^{Cr_7C_3} \tag{2.1}$$

$$n_{Al}^{tot,a.o} = n_{Al}^{Cr_2AlC,a.o} + n_{Al}^{oxide} + n_{Al}^{ppt,Cr_2AlC} + n_{Al}^{ppt,Cr_7C_3} \tag{2.2}$$

$n_{Cr}^{Cr_2AlC,a.o}$, $n_{Cr}^{oxide}$, and $n_{Cr}^{Cr_7C_3}$ denote the number of moles of Cr present in the $Cr_2AlC$ remaining after oxidation, in the oxide scale, and in $Cr_7C_3$, respectively.

$n_{Al}^{Cr_2AlC,a.o}$, $n_{Al}^{oxide}$, $n_{Al}^{ppt,Cr_2AlC}$, and $n_{Al}^{ppt,Cr_7C_3}$ denote the number of moles of Al present in the Cr$_2$AlC remaining after oxidation, in the oxide scale, in the Al-O-C-N precipitates embedded in the undecomposed MAX phase region, and in the Al-O-C-N precipitates embedded in the carbide, respectively.

Since the total number of Cr and Al atoms are conserved, as it is assumed that there is no volatilization, $n_{Cr}^{tot,a.o}:n_{Al}^{tot,a.o} = n_{Cr}^{tot,b.o}:n_{Al}^{tot,b.o} = 2$. Even though trace amounts of Cr$_2$Al were observed in equiaxed coatings after oxidation, the overall Cr:Al stoichiometry remains unchanged, as Cr$_2$Al possesses the same Cr:Al ratio as Cr$_2$AlC.

We assume that no volatile species containing Cr or Al is formed during oxidation. Therefore,

$$\frac{n_{Cr}^{tot,b.o}}{n_{Al}^{tot,b.o}} = \frac{n_{Cr}^{tot,a.o}}{n_{Al}^{tot,a.o}} = 2 \tag{2.3}$$

Dividing (2.2) by (2.1) gives,

$$\frac{n_{Cr}^{tot,a.o}}{n_{Al}^{tot,a.o}} = \frac{n_{Cr}^{Cr_2AlC,a.o} + n_{Cr}^{oxide} + n_{Cr}^{Cr_7C_3}}{n_{Al}^{Cr_2AlC,a.o} + n_{Al}^{oxide} + n_{Al}^{ppt,Cr_2AlC} + n_{Al}^{ppt,Cr_7C_3}} \tag{2.4}$$

Substituting (2.3) in (2.4),

$$2 = \frac{n_{Cr}^{Cr_2AlC,a.o} + n_{Cr}^{oxide} + n_{Cr}^{Cr_7C_3}}{n_{Al}^{Cr_2AlC,a.o} + n_{Al}^{oxide} + n_{Al}^{ppt,Cr_2AlC} + n_{Al}^{ppt,Cr_7C_3}}$$

Expanding this expression gives,

$$2\left(n_{Al}^{Cr_2AlC,a.o} + n_{Al}^{oxide} + n_{Al}^{ppt,Cr_2AlC} + n_{Al}^{ppt,Cr_7C_3}\right) = n_{Cr}^{Cr_2AlC,a.o} + n_{Cr}^{oxide} + n_{Cr}^{Cr_7C_3} \tag{2.5}$$

When there is no partial de-intercalation of Al, and all Al de-intercalation is assumed to contribute to carbide formation, the remaining Cr$_2$AlC post-oxidation is expected to retain the initial stoichiometry of Cr and Al, which gives,

$$\frac{n_{Cr}^{Cr_2AlC,a.o}}{n_{Al}^{Cr_2AlC,a.o}} = 2 \tag{2.6}$$

Substituting (2.6) in (2.5) gives,

$$2n_{Al}^{oxide} + 2n_{Al}^{ppt,Cr_2AlC} + 2n_{Al}^{ppt,Cr_7C_3} = n_{Cr}^{oxide} + n_{Cr}^{Cr_7C_3} \tag{2.7}$$

If $n_{Cr_7C_3}$ denotes the number of Cr₇C₃ moles, Equation 2.7 becomes,

$$2n_{Al}^{oxide} + 2n_{Al}^{ppt,Cr_2AlC} + 2n_{Al}^{ppt,Cr_7C_3} = n_{Cr}^{oxide} + 7n_{Cr_7C_3} \tag{2.8}$$

Equation 2.8 was used to estimate the volume of Cr₇C₃ expected to form upon oxidation. The remaining variables in Equation 2.8 were calculated using a combination of data from SEM-tomography, STEM-EDX, and APT, as explained in the following text.

(a) $n_{Al}^{oxide}$ and $n_{Cr}^{oxide}$

The total number of oxide moles $n^{oxide}$ can be related to its mass $m^{oxide}$ and its molar mass $M^{oxide}$ by Equation 2.8a.

$$n^{oxide} = \frac{m^{oxide}}{M^{oxide}} \tag{2.8a}$$

$m^{oxide}$ can be calculated using the oxide volume $V^{oxide}$, which is measured by tomography and the oxide density $\rho^{oxide}$

$$m^{oxide} = V^{oxide} \cdot \rho^{oxide} \tag{2.8b}$$

$\rho^{oxide}$ and $M^{oxide}$ are calculated considering the theoretical densities and molar masses of the oxide phases identified by TKD and SAED in our previous work [37]. For the case of the oxide forming on the equiaxed coating, the presence of Cr in the oxide, as measured by STEM-EDX, is also considered.

If the oxide contains $x$ at.% Cr, $n_{Al}^{oxide}$ and $n_{Cr}^{oxide}$, can be obtained from $n^{oxide}$ as,

$$n_{Al}^{oxide} = 2 \cdot \frac{40-x}{40} \cdot n^{oxide} \tag{2.8c}$$

$$n_{Cr}^{oxide} = 2 \cdot \frac{x}{40} \cdot n^{oxide} \tag{2.8d}$$

(b) $n_{Al}^{ppt,Cr_2AlC}$

The Al in Al-O-C-N precipitates embedded in the undecomposed Cr$_2$AlC increases the overall Al composition measured in the undecomposed region. Given that the volume percent of these precipitates is determined by APT to be only 1.6 % and they are too small to be resolved by SEM, we treat that the entire undecomposed region, including the embedded precipitates, as Cr$_2$AlC for this analysis. Under this assumption, the number of excess Al atoms present in the undecomposed Cr$_2$AlC ($N_{Al}^{ppt,Cr_2AlC}$) can be calculated from the measured excess atomic fraction Al $\Delta x_{Al}$ using Equations 2.8e and 2.8f.

We consider the number of atoms in the undecomposed Cr$_2$AlC are denoted by $N_{tot}^{Cr_2AlC,a.o.}$, the tomography-obtained volume of the undecomposed Cr$_2$AlC as $V^{Cr_2AlC,a.o.}$, the theoretical density of Cr$_2$AlC as $\rho^{Cr_2AlC.}$, the molar mass of Cr$_2$AlC as $M^{Cr_2AlC.}$ and Avogadro's constant as N$_A$.

$$N_{tot}^{Cr_2AlC,a.o.} = \frac{V^{Cr_2AlC,a.o.} \cdot \rho^{Cr_2AlC.}}{M^{Cr_2AlC.}} \cdot N_A \cdot 5 \tag{2.8e}$$

$$N_{Al}^{ppt,Cr_2AlC} = \Delta x_{Al} \cdot N_{tot}^{Cr_2AlC,a.o.} \tag{2.8f}$$

$$n_{Al}^{ppt,Cr_2AlC} = \frac{N_{Al}^{ppt,Cr_2AlC}}{N_A} = \Delta x_{Al} \cdot \frac{V^{Cr_2AlC,a.o.} \cdot \rho^{Cr_2AlC.}}{M^{Cr_2AlC.}} \cdot 5 \tag{2.8g}$$

We note that even slight changes in $\Delta x_{Al}$ have a significant effect on the estimated $n_{Cr_7C_3}$. Therefore, the $\Delta x_{Al}$ obtained from APT measurements were used, as this technique allows us to resolve the MAX phase from the precipitate. Moreover, the method benefits from the minimization of systematic errors, resulting in the smallest overall uncertainty.

(c) $n_{Al}^{ppt,Cr_7C_3}$

The carbide regions contain, say, $x$ at. % of Al in the form of Al-O-C-N precipitates. As a formula unit of $Cr_7C_3$ contains 10 atoms, each mole of $Cr_7C_3$ there will be $10.x$ moles of Al atoms present as Al-O-C-N precipitates. This gives:

$$n_{Al}^{ppt,Cr_7C_3} = 10.x.n_{Cr_7C_3} \qquad (2.8h)$$

The amount of Al present in the Al-O-C-N precipitates in the carbides was estimated using STEM-EDX, with absorption correction performed for the density of $Cr_7C_3$. Substituting Equation 2.8h in 2.8, we get

$$n_{Cr_7C_3} = \frac{2n_{Al}^{oxide} + 2n_{Al}^{ppt,Cr_2AlC} - n_{Cr}^{oxide}}{7 - 0.2x} \qquad (2.9)$$

Substituting expressions from 2.8c, 2.8d and 2.8g and using the tomography obtained volumes, theoretical densities and molar masses, and the compositions obtained from APT and STEM-EDX outlined above, $n_{Cr_7C_3}$ can be estimated.

The volume of $Cr_7C_3$ expected to form $V_{Cr_7C_3}^{est}$ can therefore be estimated using the molar volume of $Cr_7C_3$, $V_{m,Cr_7C_3}$, as follows

$$V_{Cr_7C_3}^{est} = n_{Cr_7C_3}.V_{m,Cr_7C_3} \qquad (3)$$

$V_{Cr7C3}$ is the molar volume of $Cr_7C_3$

**Acknowledgments**

The authors gratefully acknowledge the GRS funding agency [grant no. 1501626] for the support. D.J.R. is grateful to colleagues at Materials Chemistry, especially Dr. Stanislav Mráz, for valuable discussions.

**Contributions**

D.J.R conceived the research with inputs from J.M.S. D.J.R performed the experiments and analysis, theoretical calculations and wrote the original draft of the manuscript. S.A.S synthesized the coatings, contributed to valuable discussions, reviewed and edited the manuscript. J.M.S acquired funding and resources, supervised the work, reviewed and edited the manuscript.

**Competing interests**

There are no conflicts of interest to declare.

**Data availability**

Data used in the manuscript will be made available on request.

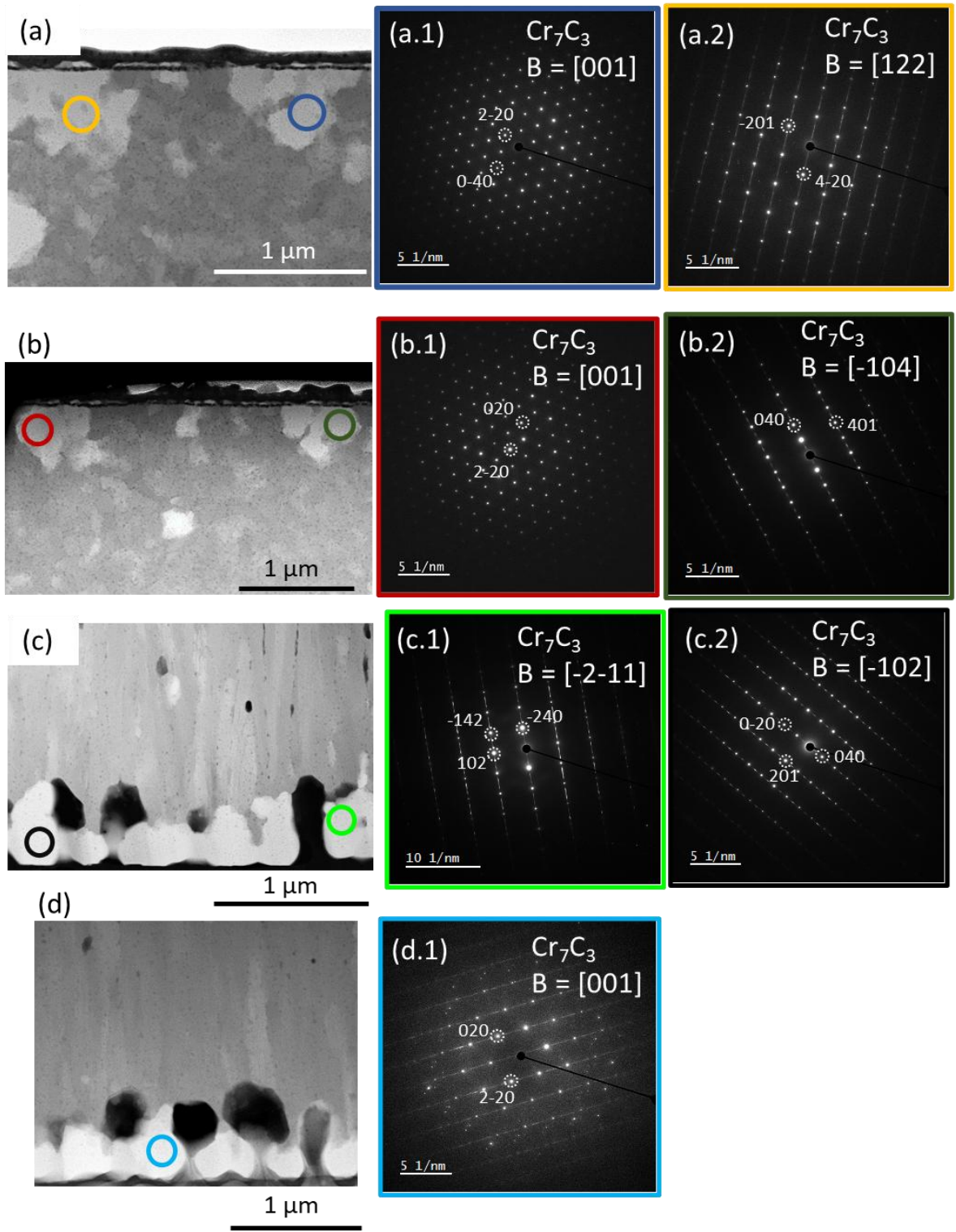

*Figure S. 1 HAADF - STEM images of (a), (b) equiaxed coatings and (c), (d) columnar coatings oxidized up to 990 °C from 2 different lamellae each. (a.1) - (d.1) correspond to the SAED patterns of the regions indicated in the images (a) – (d), respectively.*

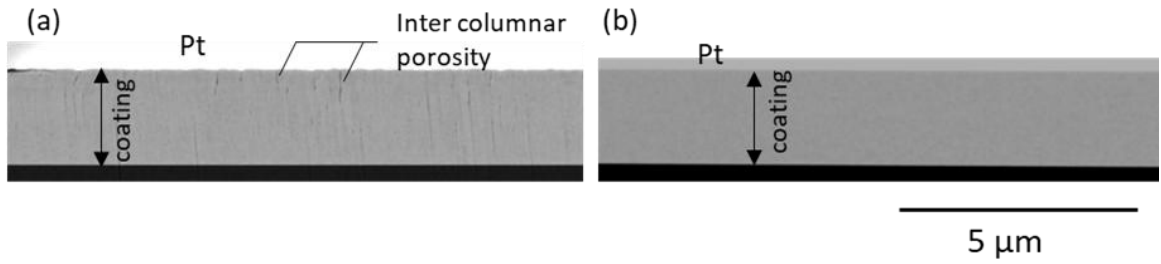

*Figure S. 2 Cross-sectional SEM images of as-synthesized (a) columnar and (b) equiaxed coatings*

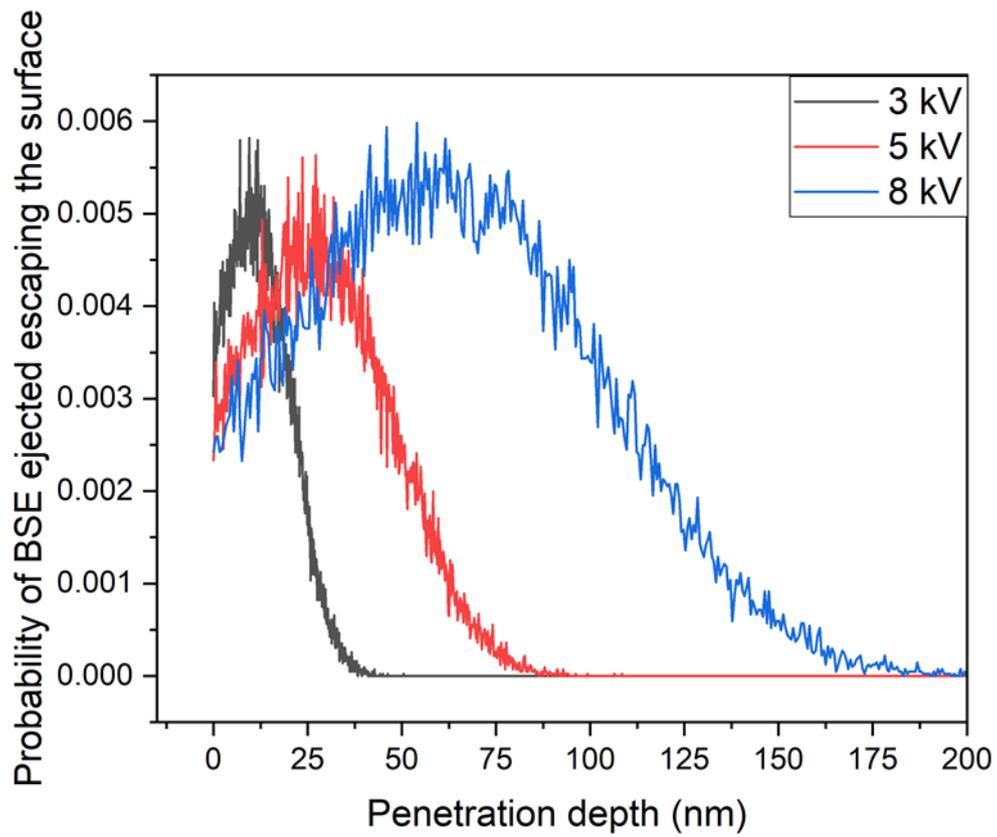

*Figure S. 3 Probability of back scattered electrons (BSE) being ejected as a function of the penetration depth at different voltages, calculated by Monte Carlo simulations*

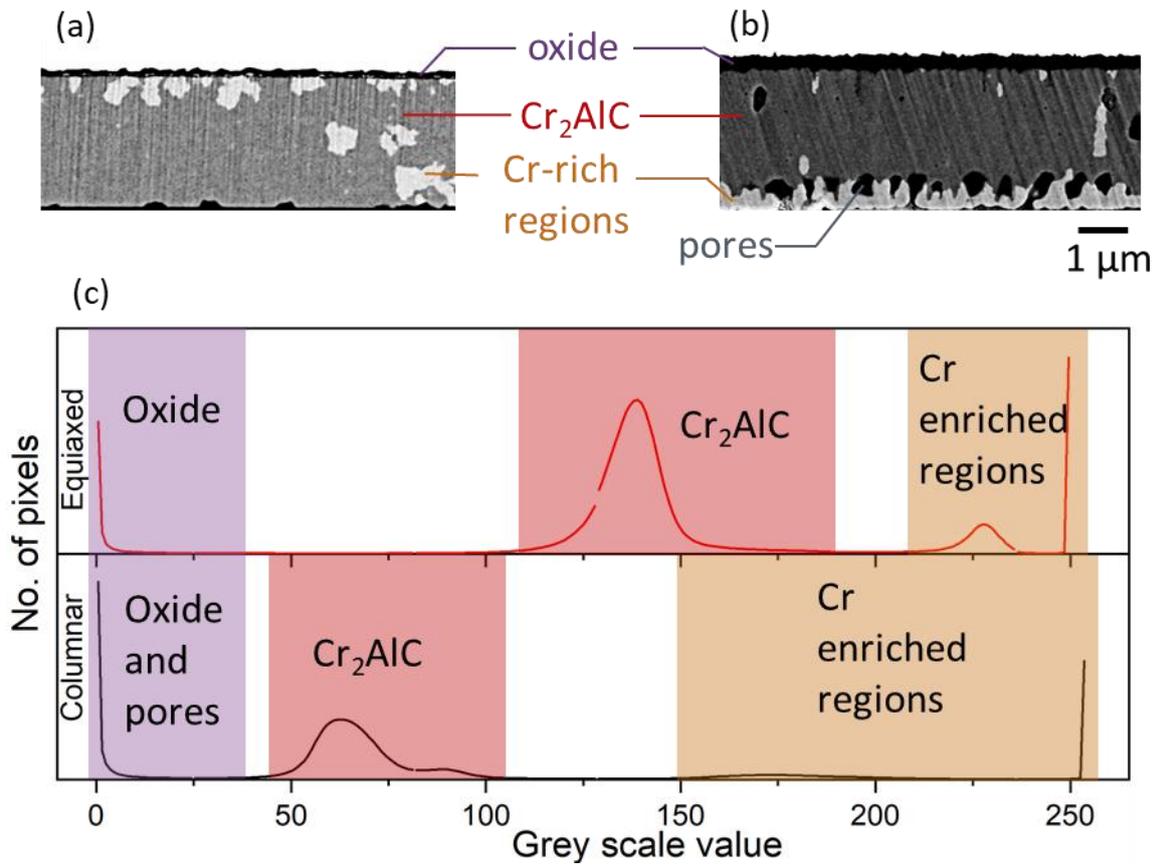

*Figure S. 4 (a) and (b) Representative SEM BSE images of equiaxed and columnar coatings oxidized up to 990 °C, respectively. (c) Plot indicating the number of pixels vs. the grey scale values for the 3-D image stack of coatings oxidized up to 990 °C.*

Table S. 1 O and N contents measured by STEM-EDX in the $Cr_2AlC$ regions at different conditions. The elevated oxygen content observed by STEM-EDX compared to ERDA and EBS measurements can be attributed to the interaction of the lamella with the atmospheric oxygen. Due to the limited MAX phase region remaining after 3 hours of oxidation, only one mapping region was available, preventing the estimation of an error bar.

| Sample condition | Morphology | O (at. %) | N (at. %) |
|---|---|---|---|
| As-synthesized | Equiaxed | 3.1 ± 0.5 | 1.5 ± 0.1 |
|  | Columnar | 3.3 ± 0.2 | - |
| Up to 990 °C | Equiaxed | 3.3 ± 0.4 | 1.6 ± 0.1 |
|  | Columnar | 3.2 ± 0.6 | - |
| 2h at 990 °C | Equiaxed | 3.9 ± 0.2 | 1.5 ± 0.2 |
|  | Columnar | 3.6 ± 1 | - |
| 3h at 990 °C | Equiaxed | 3.7 ± 0.2 | 1.6 ± 0.2 |
|  | Columnar | 4.1 | - |

# References

1. Schmitt, G., Schütze, M., Hays, G. F., Burns, W., Han, E. H., & Pourbaix, A. Global needs for knowledge dissemination, research, and development in materials deterioration and corrosion control. *World Corrosion Organization, 38,* **14** (2009).

2. Mr. Ali Kosari *et al.* Application of in-situ liquid cell transmission electron microscopy in corrosion studies: a critical review of challenges and achievements.

3. Bulavchenko, O. A. & Vinokurov, Z. S. In Situ X-ray Diffraction as a Basic Tool to Study Oxide and Metal Oxide Catalysts. *Catalysts* **13,** 1421; 10.3390/catal13111421 (2023).

4. Michael D. Uchic, Lorenz Holzer, Beverley J. Inkson, Edward L. Principe, and Paul Munroe. Three-Dimensional Microstructural Characterization Using Focused Ion Beam Tomography. *MRS Bulletin* **32,** 408–416 (2007).

5. Michael D. Uchic, Michael A. Groeber, and Anthony D. Rollett. Automated Serial sectioning methods for rapid collection of 3-D microstructure data (2011).

6. Burnett, T. L. *et al.* Large volume serial section tomography by Xe Plasma FIB dual beam microscopy. *Ultramicroscopy* **161,** 119–129; 10.1016/j.ultramic.2015.11.001 (2016).

7. Echlin, M. P., Burnett, T. L., Polonsky, A. T., Pollock, T. M. & Withers, P. J. Serial sectioning in the SEM for three dimensional materials science. *Current Opinion in Solid State and Materials Science* **24,** 100817; 10.1016/j.cossms.2020.100817 (2020).

8. Groeber, M. A., Haley, B. K., Uchic, M. D., Dimiduk, D. M. & Ghosh, S. 3D reconstruction and characterization of polycrystalline microstructures using a FIB–SEM system. *Materials Characterization* **57,** 259–273; 10.1016/j.matchar.2006.01.019 (2006).

9.  West, G. D. & Thomson, R. C. Combined EBSD/EDS tomography in a dual-beam FIB/FEG-SEM. *Journal of microscopy* **233,** 442–450; 10.1111/j.1365-2818.2009.03138.x (2009).

10. Schaffer, M., Wagner, J., Schaffer, B., Schmied, M. & Mulders, H. Automated three-dimensional X-ray analysis using a dual-beam FIB. *Ultramicroscopy* **107,** 587–597; 10.1016/j.ultramic.2006.11.007 (2007).

11. Hutzenlaub, T., Thiele, S., Zengerle, R. & Ziegler, C. Three-Dimensional Reconstruction of a LiCoO2 Li-Ion Battery Cathode. *Electrochem. Solid-State Lett.* **15,** A33; 10.1149/2.002203esl (2012).

12. Kanno, D., Shikazono, N., Takagi, N., Matsuzaki, K. & Kasagi, N. Evaluation of SOFC anode polarization simulation using three-dimensional microstructures reconstructed by FIB tomography. *Electrochimica Acta* **56,** 4015–4021; 10.1016/j.electacta.2011.02.010 (2011).

13. Liu, Z. *et al.* Three-dimensional morphological measurements of LiCoO2 and LiCoO2/Li(Ni1/3Mn1/3Co1/3)O2 lithium-ion battery cathodes. *Journal of Power Sources* **227,** 267–274; 10.1016/j.jpowsour.2012.11.043 (2013).

14. Etiemble, A. *et al.* Evolution of the 3D Microstructure of a Si-Based Electrode for Li-Ion Batteries Investigated by FIB/SEM Tomography. *J. Electrochem. Soc.* **163,** A1550-A1559; 10.1149/2.0421608jes (2016).

15. Zekri, A., Knipper, M., Parisi, J. & Plaggenborg, T. Microstructure degradation of Ni/CGO anodes for solid oxide fuel cells after long operation time using 3D reconstructions by FIB tomography. *Physical chemistry chemical physics : PCCP* **19,** 13767–13777; 10.1039/c7cp02186k (2017).

16. Cempura, G. & Kruk, A. Microstructural analysis of Sanicro 25 (42Fe22Cr25NiWCuNbN) after oxidation in steam for 25,000 h at 700 °C. *Materials Characterization,* 114751; 10.1016/j.matchar.2025.114751 (2025).

17. Coghlan, L., Shin, A., Pearson, J., Jepson, M. A. E. & Higginson, R. L. Using a plasma FIB system to characterise the porosity through the oxide scale formed on 9Cr-1Mo steel exposed to CO2. *J Mater Sci* **57,** 17849–17869; 10.1007/s10853-022-07758-9 (2022).

18. Wang, Z. *et al.* High-performance Cr2AlC MAX phase coatings: Oxidation mechanisms in the 900–1100°C temperature range. *Corrosion Science* **167,** 108492; 10.1016/j.corsci.2020.108492 (2020).

19. Michaël Ougier, Alexandre Michau, Fernando Lomello, Frederic Schuster, Hicham Maskrot, Michel L. Schlegel. High-temperature oxidation behavior of HiPIMS as-deposited Cr–Al–C and annealed Cr2AlC coatings on Zr-based alloy. *Journal of Nuclear Materials* (2019).

20. Li, Z. *et al.* High-performance Cr2AlC MAX phase coatings for ATF application: interface design and oxidation mechanism. *Corrosion Communications*; 10.1016/j.corcom.2023.10.001 (2024).

21. Li, Y. *et al.* Excellent oxidation resistance of rapidly crystallizing Cr2AlC coatings at high temperatures: Effect of microstructure. *Journal of the European Ceramic Society* **44,** 6343–6355; 10.1016/j.jeurceramsoc.2024.04.027 (2024).

22. Tang, C. *et al.* High-temperature oxidation and hydrothermal corrosion of textured Cr2AlC-based coatings on zirconium alloy fuel cladding. *Surface and Coatings Technology* **419,** 127263; 10.1016/j.surfcoat.2021.127263 (2021).

23. Imtyazuddin, M., Mir, A. H., Tunes, M. A. & Vishnyakov, V. M. Radiation resistance and mechanical properties of magnetron-sputtered Cr2AlC thin films. *Journal of Nuclear Materials* **526,** 151742; 10.1016/j.jnucmat.2019.151742 (2019).

24. Ward, J. *et al.* Corrosion performance of Ti3SiC2, Ti3AlC2, Ti2AlC and Cr2AlC MAX phases in simulated primary water conditions. *Corrosion Science* **139,** 444–453; 10.1016/j.corsci.2018.04.034 (2018).

25. Gonzalez-Julian, J., Mauer, G., Sebold, D., Mack, D. E. & Vassen, R. Cr 2 AlC MAX phase as bond coat for thermal barrier coatings: Processing, testing under thermal gradient loading, and future challenges. *J Am Ceram Soc* **103,** 2362–2375; 10.1111/jace.16935 (2020).

26. Gonzalez-Julian, J., Go, T., Mack, D. E. & Vaßen, R. Thermal cycling testing of TBCs on Cr2AlC MAX phase substrates. *Surface and Coatings Technology* **340,** 17–24; 10.1016/j.surfcoat.2018.02.035 (2018).

27. Shamsipoor, A., Mousavi, B., Razavi, M., Bahamirian, M. & Farvizi, M. Cr2AlC MAX phase: A promising bond coat TBC material with high resistance to high temperature oxidation. *Ceramics International* **51,** 6439–6447; 10.1016/j.ceramint.2024.12.088 (2025).

28. Lee, D. B., Nguyen, T. D., Han, J. H. & Park, S. W. Oxidation of Cr2AlC at 1300°C in air. *Corrosion Science* **49,** 3926–3934; 10.1016/j.corsci.2007.03.044 (2007).

29. Hajas, D. E. *et al.* Oxidation of Cr2AlC coatings in the temperature range of 1230 to 1410°C. *Surface and Coatings Technology* **206,** 591–598; 10.1016/j.surfcoat.2011.03.086 (2011).

30. Azina, C. *et al.* Microstructural and compositional design of Cr2AlC MAX phases and their impact on oxidation resistance. *Journal of the European Ceramic Society* **44,** 4895–4904; 10.1016/j.jeurceramsoc.2024.02.037 (2024).

31. Zuber, A. *et al.* Towards a better understanding of the high-temperature oxidation of MAX phase Cr2AlC. *Journal of the European Ceramic Society* **42,** 2089–2096; 10.1016/j.jeurceramsoc.2021.12.057 (2022).

32. Lee, D. B. & Park, S. W. Oxidation of Cr2AlC Between 900 and 1200 °C in Air. *Oxid Met* **68,** 211–222; 10.1007/s11085-007-9071-0 (2007).

33. Lee, D. B. & Nguyen, T. D. Cyclic oxidation of Cr2AlC between 1000 and 1300°C in air. *Journal of Alloys and Compounds* **464,** 434–439; 10.1016/j.jallcom.2007.10.018 (2008).

34. Tian, W., Wang, P., Kan, Y. & Zhang, G. Oxidation behavior of Cr2AlC ceramics at 1,100 and 1,250 °C. *J Mater Sci* **43,** 2785–2791; 10.1007/s10853-008-2516-2 (2008).

35. Reuban, A. *et al.* Unveiling the diffusion pathways under high-temperature oxidation of Cr2AlC MAX phase via nanoscale analysis. *Corrosion Science,* 112179; 10.1016/j.corsci.2024.112179 (2024).

36. Chen, X. *et al.* Enhancing the high temperature oxidation behavior of Cr 2 AlC coatings by reducing grain boundary nanoporosity. *Materials Research Letters* **9,** 127–133; 10.1080/21663831.2020.1854358 (2021).

37. Ramesh, D. J., Hans, M., Salman, S. A., Lellig, S., Primetzhofer, D., Michler, J., & Schneider, J. M. Comparative Oxidation Behavior of Columnar and Equiaxed Cr2AlC Coatings. *Corrosion Science* **253,** 112992 (2025).

38. Goldstein, J. I., Newbury, D. E., Michael, J. R., Ritchie, N. W., Scott, J. H. J., & Joy, D. C. Scanning electron microscopy and X-ray microanalysis. *Springer* (2017).

39. Pennycook, S. J. & Nellist, P. D. *Scanning Transmission Electron Microscopy* (Springer New York, New York, NY, 2011).

40. Tian, W. *et al.* Synthesis and characterization of Cr2AlC ceramics prepared by spark plasma sintering. *Materials Letters* **61,** 4442–4445; 10.1016/j.matlet.2007.02.023 (2007).

41. Azina, C. *et al.* Formation of 3D Cr2C through solid state reaction-mediated Al extraction within Cr2AlC/Cu thin films. *Nanoscale* **17,** 5447–5455; 10.1039/d4nr03664f (2025).

42. Lughi, V., Tolpygo, V. K. & Clarke, D. R. Microstructural aspects of the sintering of thermal barrier coatings. *Materials Science and Engineering: A* **368,** 212–221; 10.1016/j.msea.2003.11.018 (2004).

43. Lin, Z. J., Li, M. S., Wang, J. Y. & Zhou, Y. C. High-temperature oxidation and hot corrosion of Cr2AlC. *Acta Materialia* **55,** 6182–6191; 10.1016/j.actamat.2007.07.024 (2007).

44. Petrov, I., Barna, P. B., Hultman, L. & Greene, J. E. Microstructural evolution during film growth. *Journal of Vacuum Science & Technology A: Vacuum, Surfaces, and Films* **21,** S117-S128; 10.1116/1.1601610 (2003).

45. Chen, Y. T., Music, D., Shang, L., Mayer, J. & Schneider, J. M. Nanometre-scale 3D defects in Cr2AlC thin films. *Scientific reports* **7,** 984; 10.1038/s41598-017-01196-3 (2017).

46. Drouin, Dominique, Alexandre Réal Couture, Dany Joly, Xavier Tastet, Vincent Aimez, and Raynald Gauvin. CASINO V2.42—A Fast and Easy-to-use Modeling Tool for Scanning Electron Microscopy and Microanalysis Users _ 10.1002_sca.20000_Science Hub. *Scanning: The Journal of Scanning Microscopies* (2007).

47. Fager, C. *et al.* Optimization of FIB-SEM Tomography and Reconstruction for Soft, Porous, and Poorly Conducting Materials. *Microscopy and microanalysis* **26,** 837–845; 10.1017/S1431927620001592 (2020).

48. Demers, H. *et al.* Three-dimensional electron microscopy simulation with the CASINO Monte Carlo software. *Scanning* **33,** 135–146; 10.1002/sca.20262 (2011).

49. Tirumalasetty, G. K. *et al.* Structural tale of two novel (Cr, Mn)C carbides in steel. *Acta Materialia* **78,** 161–172; 10.1016/j.actamat.2014.06.035 (2014).

50. Hill, A. H., Harrison, A., Dickinson, C., Zhou, W. & Kockelmann, W. Crystallographic and magnetic studies of mesoporous eskolaite. *Microporous and Mesoporous Materials* **130,** 280–286; 10.1016/j.micromeso.2009.11.021 (2010).

51. Electric field gradients and charge density in corundum, α-Al2O3. Lewis, J., D. Schwarzenbach, and H. D. Flack. *Foundations of Crystallography* **38** (1982).

52. Rundqvist, Stig, and G. J. A. C. S. Runnjso. Crystal structure refinement of Cr 3 C 2. *Acta Chem Scand* **23** (1969).

53. Zhou, R.-S. & Snyder, R. L. Structures and transformation mechanisms of the [eta], [gamma] and [theta] transition aluminas.

54. Jeitschko, W., H. Nowotny, and F. Benesovsky. Kohlenstoffhaltige ternäre Verbindungen (H-phase). *Monatshefte für Chemie und verwandte Teile anderer Wissenschaften* (1963).